\documentclass[preprint,preprintnumbers,amsmath,amssymb]{revtex4}

\usepackage{graphicx}
\usepackage{bm}

\begin{document}

\title{Spontaneous Skyrmion Ground States in Magnetic Metals}

\author{U.K. R\"o\ss ler}
\affiliation{
IFW Dresden,
P.O. Box 270116,
D--01171 Dresden, Germany
}
\author{A.N.\ Bogdanov}
\altaffiliation{Permanent address: 
Donetsk Institute for Physics and Technology,
340114 Donetsk, Ukraine}
%
%
%
\affiliation{
IFW Dresden,
P.O. Box 270116,
D--01171 Dresden, Germany
}
\author{C. Pfleiderer}
\affiliation{
Physik Department E21, 
Technische Universit{\"a}t M{\"u}nchen,\\ 
James-Franck-Strasse, 
D-85748 Garching, Germany}
%
%

\date{\today}

\maketitle


\textbf{
Since the 1950s Heisenberg and others have attempted 
to explain the appearance of countable particles 
in quantum field theory in terms of
stable localized field configurations \cite{heis67}.
As an exception Skyrme's model succeeded 
to describe nuclear particles as localized states,
so-called 'skyrmions', within a non-linear field theory \cite{skyr61}.
Skyrmions are a characteristic of 
non-linear continuum models 
ranging from microscopic to cosmological scales 
\cite{durr02,baeu96,brey95,alkh01}.
Skyrmionic states have been found 
under non-equilibrium conditions, 
or when stabilised by external fields
or the proliferation of topological defects.
Examples are Turing patterns in classical liquids \cite{cros93},
spin textures in quantum Hall magnets \cite{sond93},
or the blue phases in liquid crystals \cite{wrig89}, respectively.
However, it is believed that skyrmions cannot form
spontaneous ground states 
like ferromagnetic or antiferromagnetic 
order in magnetic materials. 
Here, we show theoretically that this assumption is wrong
and that skyrmion textures may form
spontaneously in condensed matter systems with 
chiral interactions without the assistance of external fields
or the proliferation of defects.
We show this within a phenomenological
continuum model, 
that is based on a few material-specific parameters
that may be determined from experiment. 
As a new condition not considered before, we allow for softened 
amplitude variations of the magnetisation - a key property of, for instance, 
metallic magnets.
Our model implies that spontaneous skyrmion lattice ground states may 
exist quite generally in a large number of materials, notably at 
surfaces and in thin films as well as in bulk compounds,
where a lack of space inversion symmetry leads
to chiral interactions.
}

The possibility that particle-like states may form
spontaneously in continuous fields
has motivated intense theoretical efforts in the past.
Derrick and Hobart established by 
rather general arguments that particle-like configurations are 
not stable in the majority of non-linear field models 
\cite{derr64,hoba63}.
However, a few exceptions have been found.
Skyrme showed that particle-like excitations 
of continuous fields exist in the presence of certain 
non-linear interactions \cite{skyr61}.
As a drawback, the interactions considered by Skyrme
are physically not transparent, because they involve 
higher order derivative terms that are technically intractable.
Therefore, Skyrme's approach is not viable in the context 
of ordered states in condensed matter
that are ruled by short range interactions.
In contrast, a physically transparent exception 
to the Derrick-Hobart theorem has been 
recognized in systems with broken inversion symmetry,
where chiral interactions lead to skyrmion excitations 
in condensed matter systems \cite{bogd89, bogd94,bogd95}.
Chiral interactions exist in many different systems, e.g., 
(i) spin-orbit interactions in non-centrosymmetric materials,
also referred to as Dzyaloshinsky-Moriya (DM) interactions \cite{dzya64},
(ii) in non-centrosymmetric ferroelectrics,
(iii) for certain structural phase transitions,
(iv) in chiral liquid crystals, and 
(v) in the form of Chern-Simons terms in gauge field theories
(see section V in \cite{supp06} for references).
Yet, it was concluded that spontaneous
skyrmion ground states are not stable, but can 
only be induced by an appropriate applied field \cite{bogd94}
or the proliferation of topological defects  \cite{wrig89}.
Finally, a somewhat different mechanism has been found in
quantum Hall ferromagnets.
Here charge as an additional degree of freedom allows to stabilises
skyrmions in an external field.
As for the other examples skyrmions in quantum Hall ferromagnets do not
form spontaneous ground states.

We have chosen to study skyrmions that are driven
by chiral interactions, because 
the magnetism of condensed matter systems provides
perhaps the richest setting of physics problems
in which different kinds of chiral interactions may be studied. 
Present day experimental and 
theoretical studies \cite{wrig89,bogd94} have 
led to the view  that only one-dimensional 
(uniform) modulations exist as spontaneous ground states 
in these systems 
(for an illustration see Fig.\,1a).
In the following, we show that previous studies 
have been incomplete and that skyrmions may form 
spontaneously in magnets with chiral interactions in terms of 
multi-dimensional modulations, when the amplitude of the 
magnetisation is exceptionally soft.
The mathematical structure and the physical nature 
of the skyrmion texture constitute the first example 
for a spontaneous skyrmion ground state 
in an ordered condensed matter system.
The example we identify is in particular important, because
it can be tested experimentally in comparatively simple
laboratory experiments.

In our calculation we consider systems with 
a hierachy of  well-separated energy and length scales, 
where ferromagnetic exchange favouring spin alignment 
is the strongest scale, 
followed by chiral interactions favouring 
spin rotations on intermediate scales, 
and magnetic anisotropies and 
dipolar interactions that determine favourable 
directions of spins as the weakest scale.
Our model treats the magnetic states 
in chiral magnets within a 
phenomenological quasi-classical continuum approximation.
This micromagnetic approach is appropriate for
chiral magnets where the helical modulations are large 
as compared to atomic distances, i.e., 
it applies to modulations with typical lengths of a few hundred 
up to several thousand  \AA\ 
observed in many chiral helix magnets \cite{bogd89,dzya64,bak80}.
The model, hence, specifically addresses intermediate length scales.

We write the magnetic free energy density in the form
\begin{eqnarray}
 f=
A m^2 \sum_{i,j} (\partial_i n_j)^2
+\eta A(\nabla m )^2
+f_D(\mathbf{m})+f_0(m)\,,
\label{energy}
\end {eqnarray}
where the first and second term 
in $\eta $ and $A$  ($\eta,A>0$), respectively, 
describe the magnetic stiffness with $\mathbf{n}$ being
a unit vector along the magnetization, $\mathbf{m}=m\,\mathbf{n}$).
The third term $f_D(\mathbf{m}) \equiv D \mathcal{L(\mathbf{m})}$  
in the free energy describes the chiral interactions, 
where the Dzyaloshinsky-Moriya constant $D$ determines 
its handedness and strength.
The DM-interactions can be expressed
in terms of so-called Lifshitz invariants, 
i.e., linear, antisymmetric gradient terms of the magnetization 
$\mathcal{L}_{ij}^{(k)}=(m_i\partial_k m_j-m_j\partial_k m_i)$ 
where $i,j,k$ are combinations of Cartesian coordinates $x,y,z$ 
that are consistent with the symmetry class of the 
system under consideration.
The fourth term $f_0(m)$, finally, includes nongradient 
terms of the free energy and may be expanded
according to Landau theory in even powers of $m$.
We neglect weaker contributions such as magnetic 
anisotropies, or dipolar interactions and address their
importance in the discussion of our results.

The magnetic stiffness terms in $\eta$ and $A$ 
are motivated as follows.
It is convenient to consider at first local moments, 
characterised by a fixed magnetisation modulus.
For isotropic local moment ferromagnets (LMFM)
in the ordered state it is well-known \cite{akhi}, 
that the magnetisation as described in a 
continuum limit of the Heisenberg model displays 
changes of orientation \textit{and modulus}
due to magnon-magnon scattering.
This situation is described by the gradient energy given
above for $\eta=1$.
Patashinskii and Pokrovskii \cite{pata74} further
showed that magnon-magnon scattering 
in LMFM also forces
the longitudinal stiffness to vanish in the limit of infinite wavelength,
i.e., $\eta=1$ is valid in this limit also
(see also, Zwerger \cite{zwer04}). 
However, the SU(2) symmetry of the spin system enforces
that the longitudinal stiffness is the result of 
higher order transverse processes.
Further mechanisms on top of magnon-magnon scattering, 
such as coupling to the particle-hole continuum in 
itinerant-electron systems, dipolar interactions and defects, 
therefore additionally reduce the longitudinal stiffness before 
it vanishes altogether in the limit of infinite wavelength.
A minimal description of this additional 
longitudinal softening requires the second 
gradient term written in $\eta$ with $\eta<1$, 
where $\eta$ expresses the ratio 
between longitudinal and transverse 
stiffness of the ordered magnetic state.
For an extensive discussion of 
the technical validity of the two gradient terms 
used here and their relationship 
to conventional spin fluctuation theories 
of itinerant-electron magnets, we refer 
to the supplementary information of this paper \cite{supp06}.

We have also analysed \cite{supp06} published 
data of the longitudinal susceptibility 
in the local-moment ferromagnet EuS \cite{boen02} and 
the itinerant-electron ferromagnets 
Ni \cite{boen91} and MnSi \cite{sema99}.
Our analysis establishes values 
of $\eta$ for EuS, Ni and MnSi of 0.925, 0.65 and 0.4, 
respectively, i.e, important reductions of 
the longitudinal stiffness exist in all real materials.
The experimental data for these archetypical ferromagnets 
show that the regime $\eta < 1$ in our model is valid and 
relevant for many different types of magnetic systems.
In particular, the very prominent longitudinal softening 
in itinerant-electron magnets Ni and MnSi has 
not yet been appreciated before.
In the presence of chiral interactions the reductions of 
longitudinal stiffness become of tantamount importance, because they 
facilitate the formation of spontaneous skyrmion ground states.

For our model we first show that certain individual skyrmions represent
localised regions of reduced ground state energy
on a background of vanishing magnetization.
General consideration of the stability of particle-like
solutions as function of dimension against 
compression/stretching \cite{bogd95} strongly
suggest that two-dimensional solutions are the most stable.
We therefore begin by considering an individual skyrmion 
of cylindrical symmetry,
as shown in Fig.~1\,b and c to explain the basic
physical ideas.
As explained in further detail below, 
the stability of such a cylindrical structure establishes
the possibility of spontaneous skyrmion lattices 
in \textit{all} noncentrosymmetric crystallographic classes 
that allow chiral interactions \cite{bogd89}.

Expanding the contribution $f_0(m)$ 
according to Landau theory 
for small $m$ as $f_0=a(T-T_C)m^2 + bm^4+\ldots$.
leads in the absence of chiral interactions ($f_D$ = 0) 
to the conventional Curie temperature $T_C$ of centrosymmetric systems.
Thus, when the temperature drops below $T_C$ 
the energy density of a ferromagnetically spin-aligned 
state is lowest.
Chiral interactions ($f_D<0$) favouring rotations of 
the moments with respect to each other reduce the energy density further.
This stabilization of twisted magnetic states occurs 
only by the competition of the DM-interactions with the exchange,
expressed by the form of the gradient energy.
Therefore, the exact form of $f_0(m)$ is not decisive
for the stability of skyrmions.
In particular a form of $f_0(m)$ including 
higher order terms with $b<0$ and a stabilizing term $m^6$ 
for a first-order transition will change only 
quantitative relations and the details of the phase-diagram.
Possibly, it may lead to the discontinuous nucleation of skyrmions.
Figure~1 illustrates the twisted magnetic structures 
with rotating magnetization and 
the associated energy densities.
For rotations along a single given direction, 
shown in Fig.\,1\,a, the reduction in energy density, 
shown in Fig\,1\,d, is uniform.
The transition to the ferromagnetic 
state is consequently pre-empted 
by a transition to a one-dimensionally 
modulated state at a temperature
\begin{equation}
T_h=T_C + \frac{D^2}{4Aa}\,.
\label{hel-trans}
\end{equation}
As shown in Fig.\,1\,b and d rotations of the magnetisation 
in two dimensions reduce the energy density 
even further near its nucleation point at 
the center of the cylindrical structure $\rho=0$.
Considering only the energy reduction at $\rho=0$ we obtain
as an upper limit  for the temperature, $T_D$, 
for the onset of this so-called double-twisting
\begin{equation}
T_D=T_C+\frac{D^2}{2Aa}\,.
\label{skyrm-trans}
\end{equation}
The characteristic temperatures $T_h$ and $T_D$
depend only on a few physically meaningful, 
material specific parameters  
that may be determined from experiment. 
The ratio $A/D$ measures the pitch of 
the chiral modulations in the ordered helical state, 
while $D/a$ measures the chirality of magnetic fluctuations 
in the paramagnetic state. 
Hence, $D/a$ is related 
to a chiral component of the susceptibility.
The parameters $a$ and $b$ are the initial 
susceptibility and mode coupling parameter, respectively.
This permits estimates of $T_h$ and $T_D$ from experiments.
An example is given below.

As shown in Fig. 1\,d the energy reduction 
due to a double-twist is no longer uniform.
In fact, for fixed magnetisation amplitude 
the energy reduction turns 
into an excess of the energy density 
at larger distances $\rho$ from the center,
which outweighs the initial reduction.
This destabilises the double-twist structure altogether,
which is the reason why magnetic skyrmions prior to 
the work reported here were believed to be unstable.
Skyrmion textures may only form spontaneously when 
no additional energy is required for double-twist rotations 
at large $\rho$.
The latter may be achieved by permitting 
the magnitude of the moment to be soft.
As shown in Fig.\,1\,c and 1\,d skyrmions under 
these conditions have the lowest energy density, 
when the moment decreases with increasing distance 
from the centre.

The results of a comprehensive numerical 
analysis of the free energy for skyrmions 
taking into account the variation of the energy density
with $\rho$ for differing $\eta$ 
are summarised in Fig.\,2 and 3
(see Eqn.\,(\ref{energy}) and Methods section).
This analysis establishes rigorously 
that stable cylindrical skyrmions form spontaneously
in a finite temperature interval $T_t<T<T_S$
of the phase diagram Fig.\,2, where the limits
$T_h$ and $T_D$ are also marked.
The magnetic structure of the skyrmions is characterised 
by a magnetisation vector $\textbf{m}=(m,\theta,\psi)$ 
that rotates outward from the axis in all directions, 
while the modulus $m(\rho)$ shrinks and decreases towards 
the boundary of the tube at radius $R$ (Fig.~1\,c).
The amplitude of the local magnetisation  $m(\rho)$ vanishes 
continuously when approaching $T_S$ from below, as shown in Fig.~3.
Thus, the skyrmion nucleation and the formation of the 
skyrmion texture coincide.
The increase of the magnetization 
with decreasing temperature
drives the transition from 
the skyrmion phase to the uniform helix phase 
at a lower temperature $T_t$.
The transition into a one-dimensionally modulated
state by transforming the structure Fig.~1\,b or c
into the helix Fig~1\,a 
cannot be performed by a smooth deformation of the skyrmions,
because these two textures have different topology.
For this reason, a first-order transition 
is generally expected at a  temperature $T_t$.
In the phase diagram, the transition line $T_t$ 
has been determined from the approximate condition that 
the average energy of the circular skyrmion cell is equal 
to that of the helix.
The skyrmion phase vanishes 
as $T_S\to÷T_t$ in the limit $\eta\to1$.
This feature of the phase diagram (Fig.~2)
concurs with an analytical result by Wright and Mermin 
for magnets with the properties of 
fixed moment classical 
Heisenberg models (Eqn.\,(\ref{energy}) with $\eta=1$)
\cite{wrig89}.
The result was believed to prove 
the absence of skyrmion ground states 
in chiral ferromagnets.  
However, it applies only to a particular form
of the model where $\eta=1$, while the regime
$\eta<1$ had not been considered systematically.

Having established that cylindrical skyrmions
with soft amplitude can be localised regions of 
reduced energy we next address the proliferation of 
these skyrmions to form textures.
In a chiral magnet, the formation of an extended skyrmion condensate 
is subject to: (i) weak residual magnetic 
anisotropies and dipolar interactions
in the crystalline environment, and (ii) the magnetisation 
in the regime joining the skyrmion tubes.
Under the most favourable conditions, the skyrmion condensate
may form a regular lattice in analogy to the vortex lattice of 
type 2 superconductors.
In fact, there exists a fundamental analogy between magnetic 
skyrmion lattices and Abrikosov lattices.
The formation of stable spatially modulated 
states in superconductors and in magnets
can be justified by the concept of 
negative domain wall energies 
\cite{bogd89,abri57,bogd05}.
We illustrate the stability of a skyrmion condensate 
for this case by the full solution for an extended state 
in an isotropic system in two dimensions. 
We note that the solution as shown in Fig.~4
applies also to three-dimensional systems, when the solutions are artificially
constrained to remain homogeneous in the third
spatial direction (perpendicular to the plane shown in Fig.\,4).
The structure of this solution shows in particular 
that the core region of the skyrmions is unaffected, 
as stated above, while the magnetization between the skyrmions
mediating their interactions does not even need to be strictly zero
for the skyrmion lattice to become the ground state.

The spontaneous formation of skyrmions
based on cylinder-symmetric solutions 
applies to \textit{all} symmetry classes 
including higher symmetries, say the cubic 
and the isotropic case without inversion center, 
because the core structure of the skyrmions is determined by
the two largest energy scales, namely the exchange energy 
and chiral interactions,
while the texture is determined by weaker interactions.
Technically the general validity may 
be seen by noting that the Euler equations for the 
modulus of the magnetisation $m$ and the polar angle $\theta$
which control the energetic stability of the skyrmions, given by 
Eqn.\,(7) and (8) in the Methods section, respectively, 
are identical for all relevant symmetry classes 
(see Refs.\cite{bogd89,bogd95} for a discussion).
In contrast, the behaviour of the azimuthal angle $\psi$, 
which differs for the different symmetry classes, 
only affects the manner of rotation of the magnetic moments 
in the radial direction.
General considerations of the stability of localized textures 
for chiral magnets suggest that droplet-like skyrmions 
are not stable in three dimensions \cite{bogd95}. 
Therefore, we believe that the two-dimensionally localized skyrmion 
tubes discussed so far will form extended ground-state condensates 
in three-dimensional systems.
However, in a three-dimensional 
environment bending and terminating skyrmion tubes imposes a 
cost in energy that determines 
the nature of the condensate. 
The additional cost in energies may be readily reduced by 
closing the tubes amongst themselves or ending them at surface
and defects.
In isotropic three-dimensional materials
the skyrmion condensate will therefore have a liquid-like or 
amorphous appearance driven by the 
frustration between preferred orientations
and their elastic interactions in the condensed structure.

Experimental systems where skyrmion ground states may exist
will have to display softened longitudinal stiffness together 
with a hirachy of energy scales 
in which spin-aligning exchange interactions are dominant
followed by subleading chiral interactions causing spin rotations.
Further, magnetic anisotropies must be low
because they prevent the continuous rotation of 
the magnetization and, thus, multiply-modulated textures 
\cite{bogd89,bogd94,dipo}.
The elementary mechanism, shown here for a
ferromagnet and illustrated in Fig.~1, then
applies to any chiral system with a tendency 
to form twisted structures.
In particular, our model also applies to antiferromagnetic 
crystals with chiral symmetry by replacing the 
magnetization with a staggered antiferromagnetic vector order parameter
\cite{bogd02}.
Owing to the broken inversion symmetry at the surfaces
of magnetic materials \cite{fert90,bodg01}
thin films and interfaces are
the largest class of condensed matter systems 
where skyrmion condensates may form spontaneously.
Appropriate metallic films of this kind
have not been investigated in detail in the regime of interest.
A less frequent, but equally promising class 
of materials are bulk materials
with chiral interactions due to a lack of inversion symmetry.

The most powerful technique to test our prediction experimentally 
is polarised magnetic neutron scattering.
For a periodic skyrmion lattice
neutron scattering will show long-range
order akin to the vortex lattice in superconductors,
while the scattering intensity of an amorphous appearance
will be akin to partial 
order in liquid crystals \cite{chai95}.
In the bulk properties 
subtle evidence for the predicted intermediate phase is expected, 
when going from paramagnetism to helical magnetic order.
Real-space imaging of magnetic textures
by magnetic force microscopy, spin sensitive tunneling, 
magnetic x-ray microscopy, and Lorentz transmission electron microscopy 
are finally expected to provide evidence of skyrmion 
lattices akin vortex lattices in the Shubnikov phase of superconductors. 
%
%
The helical structure in the chiral magnet (Fe,Co)Si 
were recently imaged directly for the first time \cite{uchida06}
because of the exceptionally long helix period in this material,
but the technique may introduce uncontrolled strains that
interfere with the intrinsic properties.
It is to be expected that the resolution of improved imaging 
techniques will soon be sufficient to 
take pictures of the skyrmions predicted here.

Our study was inspired by recent 
high pressure experiments in the cubic itinerant-electron helimagnet MnSi.
The low temperature resistivity in MnSi changes abruptly
from a $T^2$-Fermi liquid temperature dependence to a
$T^{3/2}$ dependence 
above a pressure $p_c=14.6$\,kbar \cite{pfle01,doyr03}.
The resistivity is unchanged up to at least 3$p_c$, 
suggesting the formation of a stable phase.
However, as emphasized in Ref.\,\cite{pfle01} a 
$T^{3/2}$ resistivity is normally characteristic of spin glasses
and amorphous ferromagnets where the conduction electrons undergo a diffusive
motion, but not for the ultra-pure MnSi single-crystals investigated, 
which by all accounts should show the behaviour of a Fermi liquid.
The spontaneous formation of an amorphous magnetic state
in ultrapure samples would provide a simple explanation 
that resolves this contradiction of high purity and diffusive 
charge carrier motion in MnSi.
It also explains the stability of the resistivity as a new phase.
Magnetic neutron scattering in MnSi further revealed,
that large ordered moments are present in 
the nominally paramagnetic phase.
The ordered moments are organised such that they lead
to scattering intensity on the surface of a sphere with
unexplained broad maxima for $\langle110\rangle$. 
If we suppose that the easy magnetic axis, $\langle111\rangle$, 
does not change as function of pressure, 
while the magnetism changes from a helical 
modulation along $\langle111\rangle$
to cylindrical skyrmion tubes with their axes 
preferentially along $\langle111\rangle$ we expect
intensity in the great circles perpendicular to 
$\langle111\rangle$.
These great circles interset along the $\langle110\rangle$
directions, thus explaining the broad maxima in this location.
Finally, as shown above we predict a spontaneous skyrmion
phase prior to the onset of helical order.
From the values of $T_C=29.5$\,K, the wavelength
of the helix $q_0=0.039$~\AA$^{-1}$ and the 
effective Curie-Weiss moment $p_{eff}=2.2 \mu_B$ 
we estimate $T_D-T_C \approx 0.9$K \cite{supp06}.
A hump in the specific heat above $T_C$ and 
and an associated sphere in neutron 
scattering intensity as observed in ambient pressure
experiments in MnSi are consistent with such a 
phase over a temperature interval of $\sim1$\,K
\cite{lebe93,supp06}.
We note that recent theories on the partial magnetic order 
in MnSi by Tewari et al. \cite{tewa06} 
and Binz et al.\cite{binz06} also
suggest inhomogeneous twisted states with modulated amplitude. 
These theories rely on unsubstantiated higher-order 
interaction terms and, therefore, share 
the problems with the corresponding non-linear field theories 
without identifying the basic mechanism of stable skyrmions in 
such textures.
The evidence for a skyrmion ground state in MnSi 
discussed here resolves all of the existing contradictions, 
but a definitive proof will require a host of novel techniques
listed above.

We finally note, that the argument for the existence of skyrmion textures 
in ferromagnetic metals is closely related 
to the blue phases of chiral nematic crystals 
\cite{wrig89,horn82}.
The standard continuum model of chiral nematics
admits localized structures as vortex tubes \cite{horn82}, 
where the energy disadvantage by double-twisting a configuration
has been recognised early on \cite{wrig89}.
However, in the blue phases topological defects in the form
of disclination lines, which do not exist in magnetic systems, 
allow to overcome the disadvantage in energy
and stabilise extended textures. 
In chiral magnets skyrmion textures 
appear instead as condensate of strictly localized 
skyrmions with lines or sheets of vanishing magnetic
order between them (Figure~4).
Fluctuations and the competition betwen different 
frustrated couplings entail the possibility of 
internal transitions between various textures
with liquid, glassy, or lattice-like organization of skyrmions.
Thus, the predicted magnetic skyrmion textures promise rich phase diagrams, 
similar to those of liquid crystals and Abrikosov lattices of flux-lines in type-II superconductors
\cite{abri57}.

\section*{Acknowledgements}

We wish to thank H. Eschrig, I. Fischer, 
A. M\"obius, A. Rosch, H. v. L\"ohneysen, M. Vojta, and W. Zwerger
for support and discussions. 
We are particularly greatful to P. B\"oni for 
discussions on his studies of longitudinal magnetic fluctuations 
in EuS, Ni and MnSi .
AB thanks the DFG-Graduiertenkolleg GRK 284 'Kollektive P\"anomene im Festk\"orper' for financial support.
CP acknowledges support in the framework of 
a Helmholtz-Hochschul-Nachwuchsgruppe
at the Universit\"at Karlsruhe in the initial part of this project.
\newpage

\section*{Methods}


Cubic noncentrosymmetric magnets (e.g., MnSi and other B20 compounds
\cite{bak80}) ) belong
to  the crystallographic class $T$. For this class
the Dzyaloshinsky-Moriya energy is
\begin{equation}
f_D = D\,(\mathcal{L}_{yx}^{(z)}+\mathcal{L}_{xz}^{(y)}
+\mathcal{L}_{zy}^{(x)}) \equiv D\,\mathbf{m}\cdot \rm{rot}\, \mathbf{m}\,,
\label{DzMo}
\end{equation}
where $D$ is the Dzyaloshinsky constant \cite{bak80}.
The helix solutions for these systems are described
by a linear wave with a wave number $q_0= D/(2A)$
and constant magnetisation amplitude $m^2=2a(T_h-T)/b$
\cite{bak80,wrig89}. 
For all uniaxial systems the invariants $f_D$ 
are listed in \cite{bogd89}.
In particular, for uniaxal magnets  
with $C_{nv}$ symmetry ($n$ = 3, 4, 6)
one has
$f_D=D(\mathcal{L}_{xz}^{(x)}-\mathcal{L}_{yz}^{(y)})$.
The $n$-fold axis is directed along $z$.
Examples are Tb$_3$Al$_2$ or Dy$_3$Al$_2$ 
with tetragonal noncentrosymmetric 
structure C$_{4v}^4$ \cite{bogd94}
or the superconducting antiferromagnet CePt$_3$Si, 
space group P4$mm$ \cite{baue04}.
The structure of the cylindrically symmetric skyrmions
is given by solutions of the Euler equations
for the free energy Eqn.\,(\ref{energy}).
The magnetization vector ${\bf{m}}=(m,\theta,\psi)$
is aligned along the cylinder or $z$-axis in the skyrmion
center, $\rho=0$.
In cubic and uniaxial systems solutions 
for a skyrmion homogeneous along the $z$-direction
are given by $m (\rho)$, $\theta (\rho)$,
$\psi={\mathrm{sign}}\,\phi+\phi_{0}$,
where ${\mathrm{sign}}=+$ or $-$ and $\phi_{0}$ is a constant, 
both determined by the specific symmetry class\cite{bogd89,bogd94}.
The dependence $m (\rho)$  and $\theta (\rho)$ 
describes the variation of the amplitude of the magnetic 
moments and their angle as function of the distance $\rho$ 
to the cylinder axis in cylindrical coordinates, respectively.
After substitution of the solutions for $\psi (\phi)$
and an integration with respect to $z$ and $\phi$
the energy of the skyrmion is reduced to the following
functional 
$\mathcal{F}_s = 2\pi L \int_0^{\infty} f_s (m, \theta) \rho  d \rho$
with energy density
\begin{eqnarray}
f_s =  \eta A \left(\frac{d m}{d \rho} \right)^2
+A \mathcal{G}(\theta, \theta_{\rho}) m^2 +
q_0^{-2} f_0 (m) \,, \\
\label{energyV}
\mathcal{G}(\theta, \theta_{\rho}) =
\left(\frac{d \theta}{d \rho} \right)^2
+ \frac{\sin^2 \theta}{\rho^2}
-2 \left[  \left(\frac{d \theta}{d \rho} \right)
+\frac{\sin \theta \cos \theta}{\rho} \right]\,,
\label{energyTheta}
\end {eqnarray}
where $L$ is the length of the skyrmion along $z$-direction,
and the spatial coordinate $\rho$ is measured in $q_0^{-1}$ units.
The Euler equations for the functional (\ref{energyV})
\begin{eqnarray}
\eta \left[ \left(\frac{d^2 m}{d \rho^2} \right)
+\frac{1}{\rho}\left(\frac{d m}{d \rho} \right)  \right]
- A \mathcal{G}(\theta, \theta_{\rho})  m
-q_0^{-2}  \frac{\partial f_0}{\partial m}  = 0,
\label{equM}
\end{eqnarray}
\begin{eqnarray}
m \left[ \left(\frac{d^2 \theta}{d \rho^2} \right)
+\frac{1}{\rho} \left(\frac{d \theta}{d \rho} \right)
- \frac{\sin \theta \cos \theta}{\rho^2} - \frac{2 \sin^2 \theta}{\rho} \right]
+ 2\left[\left(  \frac{d\theta}{d \rho} \right)
- 1 \right] \left(\frac{d m}{d \rho} \right)= 0
\label{equTheta}
\end{eqnarray}
with the boundary conditions $\theta(0)$ = 0, $m(0) = m_0$, $m(R)$ = 0
describe the structure of the isolated skyrmion and yield the energy
as a function of $m_0$ and $R$. 
The equilibrium skyrmion profiles are derived 
by minimization of $\mathcal{F}_s (m_0)$ with respect to $m_0$ and $R$; the 
radially integrated energy densities of the vortices per lengths (Figure~1~d)
are given by $w(\rho)=(2/\rho^2) \int_0^{\rho} f_s (m, \theta) \rho  d \rho$.
Full solutions for extended two-dimensional models 
have been constructed numerically using 
a finite-difference scheme on rectangular lattices 
with dynamically optimized lattice parameters
and periodic boundary conditions.
The search for ground-states and checks on their stability 
were done using a continuous variable Monte Carlo simulated 
annealing method and mesh-refinement.

The unit length used throughout this paper 
is $1/q_0$, which is the inverse of the twisting length
in chiral magnets with  Dzyaloshinsky-Moriya couplings.
Energy and temperature are given 
in equivalent units of $D^2/(2Aa)$ as 
in Eqs.~(\ref{hel-trans}) and (\ref{skyrm-trans}).

\clearpage
\thispagestyle{empty}

\centerline{\includegraphics[width=16cm]{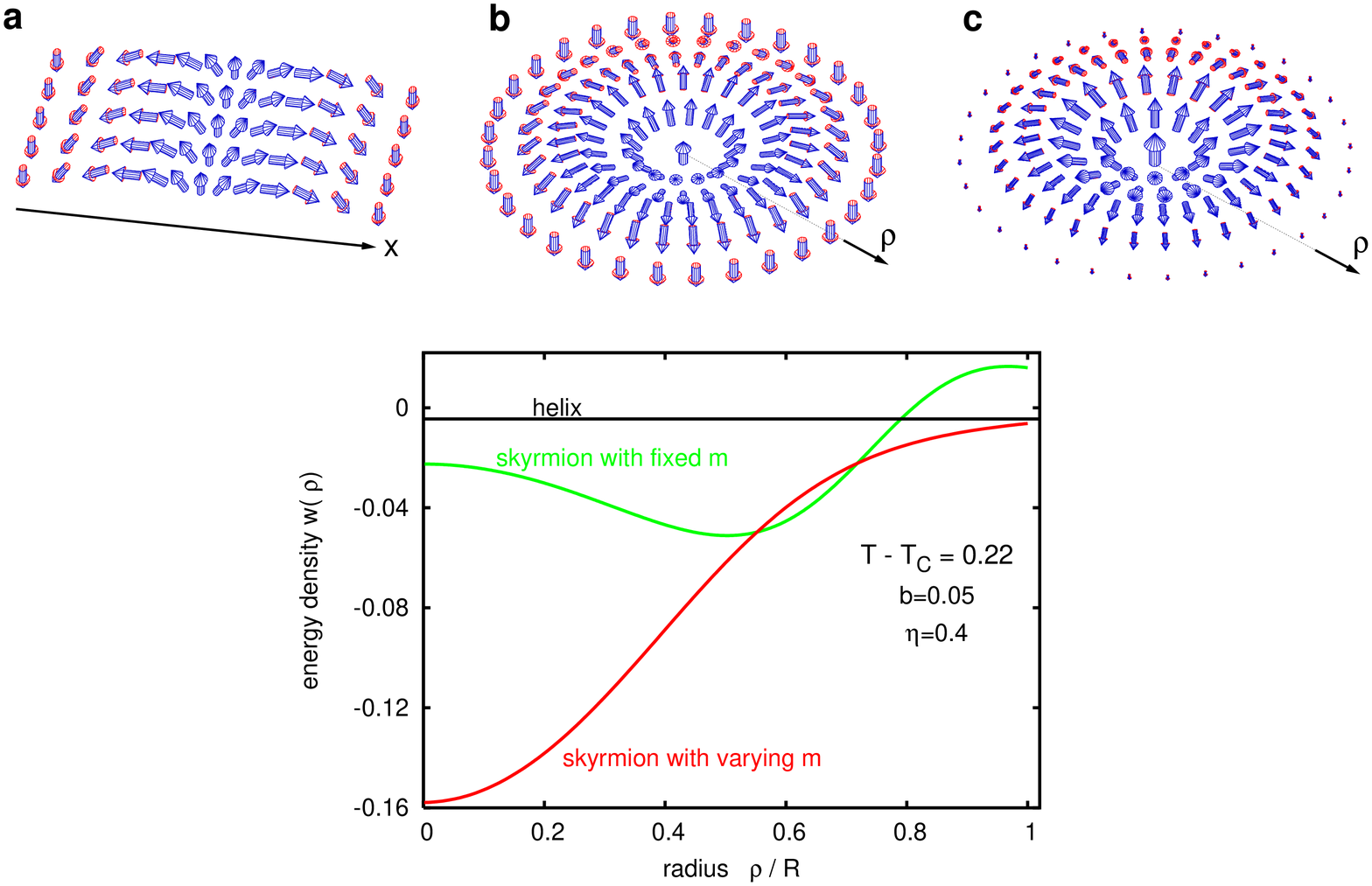}}

\bigskip
\bigskip

\Large\textbf{Fig.~1}

\clearpage

\begin{figure}[h]
\caption{
Chiral modulated structures 
for noncentrosymmetric ferromagnets 
(\textbf{a - c}).
\textbf{a}, one-dimensional modulated structure 
propagating along a direction  $x$ and
with a constant modulus of magnetic moments.
\textbf{b}, 
Circular cross section through a skyrmionic excitation 
with fixed modulus.
It can be visualized by rolling one half-period of 
a helix modulation as in \textbf{a}
into a cylindrical tube with propagation axes 
now pointing radially out in
all directions $\rho$ from the center.
\textbf{c},
A circular skyrmion cell. 
The magnetization has maximum modulus at the skyrmion axis and 
rotates along the radial directions ($\psi = \phi)$. 
The modulus $m$ gradually decreases with increasing 
distance from the center, and equals zero at the boundaries of the cell.
The chirality (handedness) of the cylindrical skyrmions shown 
here may be seen as follows: when the fingers 
on a hand follow the moments rolling outward
the thumb defines a winding sense (chirality) around the center 
(note that space inversion is not satisfied with 
respect to points in the center 
as this winding direction would have to 
be reversed upon space inversion). 
The magnetic structures \textbf{a}, \textbf{b}, \textbf{c} 
can exist in noncentrosymmetric crystals of class C$_{nv}$.
The chiral magnets Tb$_3$Al$_2$ or Dy$_3$Al$_2$ with tetragonal 
noncentrosymmetric structure C$_{4v}^4$ 
belong to this class \cite{bogd94}.
For this crystallographic class the 
magnetization rotates around an axis, that is 
perpendicular to the propagation directions in the 
one-dimensionally modulated state and tangentially 
for the cylindrical skyrmions.
In noncentrosymmetric crystals 
from the classes D$_{n}$ and in cubic or isotropic systems,
the magnetization in the helix and the skyrmions rotates 
in the planes perpendicular to the propagation directions, 
i.e., about the propgagation directions.
Such structures can exist in the noncentrosymmetric
ferromagnets with B20-structure, 
in particular in MnSi \cite{bak80}.
For other crystallographic symmetries 
more involved structures can occur.
All of these magnetic structures in three-dimensional crystals 
can be described either as one-dimensional modulated state, 
which is a helix, or as cylindrical skyrmions.
Phenomenological theory allows to analyse the stability 
of these structures for different crystallographic classes 
within a common mathematical framework, see 
Methods section and Ref.~\cite{bogd89}.
\textbf{d}, comparison for the local energy density 
per unit volume for the helical modulation, \textbf{a}, and fixed  
and varying magnetic moment $m$, 
\textbf{b} and \textbf{c}, respectively,
in circular skyrmion cells (radius $R$) 
along a radial direction $\rho$.
The corresponding helical state has a constant energy density
which is always higher than the energy density 
in the center of the skyrmions.
For fixed-moment skyrmions this energy gain is offset 
by a large excess energy 
near the boundary of the cell at $R$.
}
\label{Fig1}
\end{figure}
\clearpage
\thispagestyle{empty}

\centerline{\includegraphics[width=16cm]{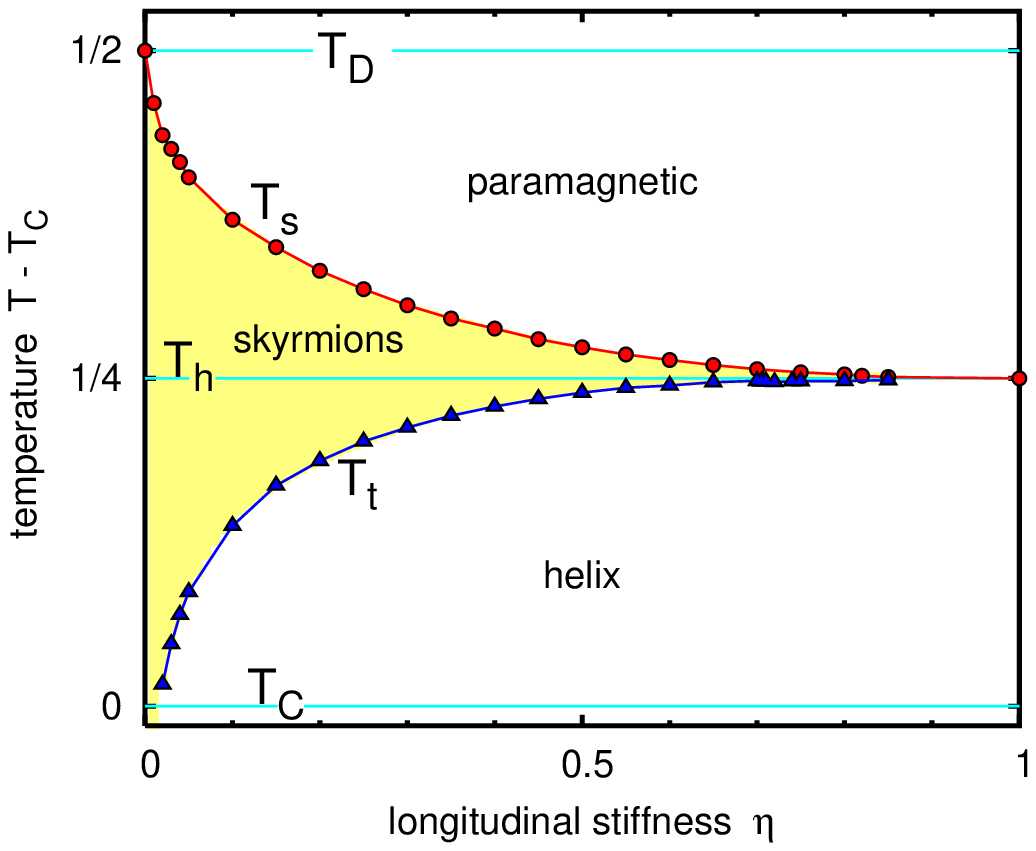}}

\bigskip
\bigskip

\Large\textbf{Fig.~2}

\clearpage

\normalsize
\begin{figure}[h]
\caption{Phase diagram of a chiral ferromagnet
in terms of temperature versus 
longitudinal stiffness-parameter  ($\eta, T$).
The diagram is based on the stability of 
the cylindrical skyrmions shown 
in Fig.~1\,c. Therefore, the diagram 
applies for two- and three-dimensions. 
Above the ferromagnetic Curie temperature $T_C$ 
two characteristic temperatures,
defined in Eqn.\,(\ref{skyrm-trans}),
rule the phase diagram:
$T_D$ for the instability against double twisting
and $T_h$ for the formation of the helical phase.
The skyrmion phase is thermodynamically stable 
between the line $T_S(\eta )$
for the continuous transition into the paramagnetic phase  
and the line $T_t(\eta)$ for the first-order transition into the helix. 
This line is independent on the chosen parameter $b$.
The helical phase exists as a metastable phase between $T_h$ and $T_t(\eta)$.
The estimate for 
the first-order line $T_t(\eta)$ 
has been calculated for models with $b=0.05$.
}
\label{Fig2}
\end{figure}

\clearpage

\centerline{\includegraphics[width=16cm]{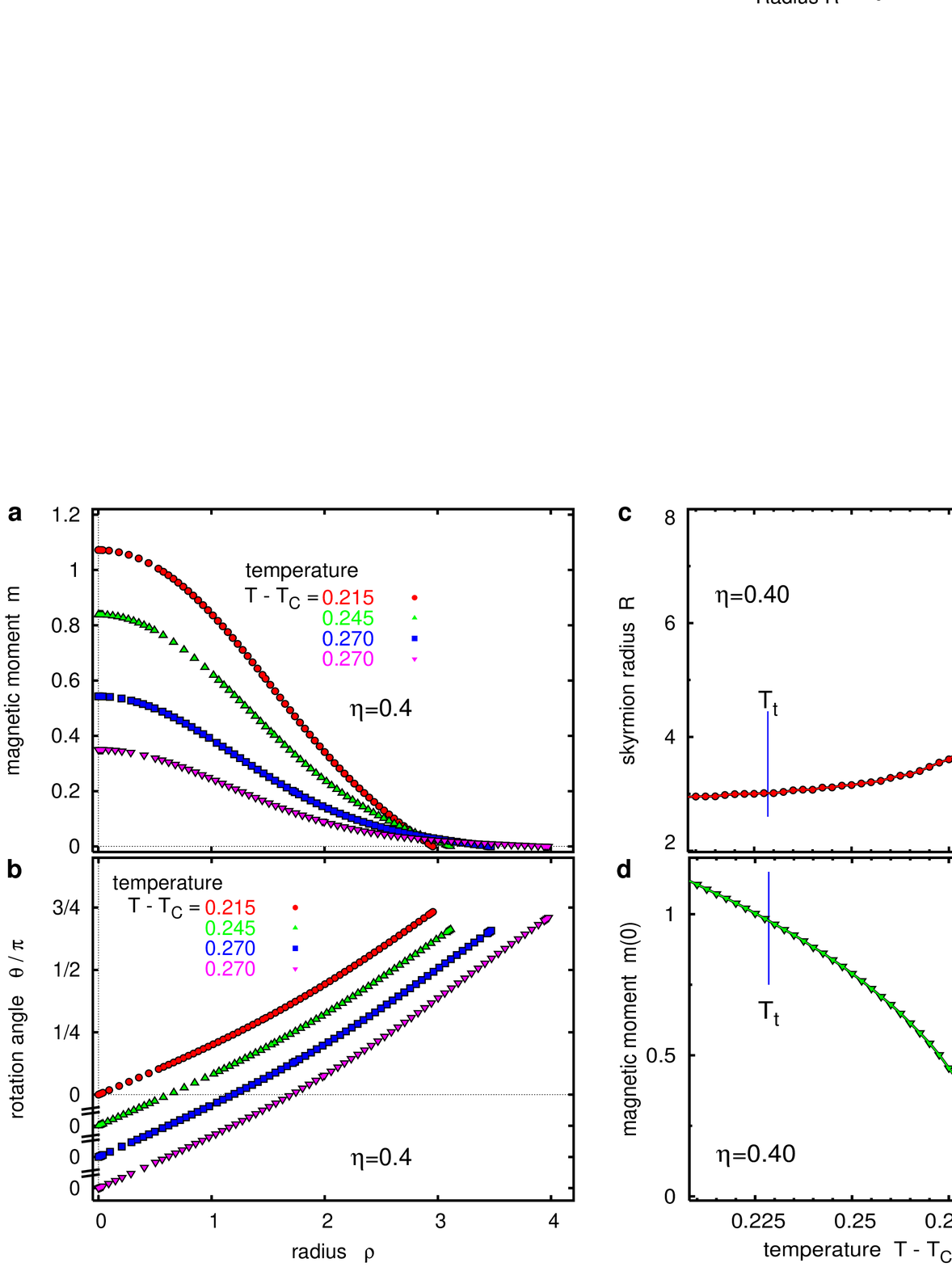}}

\bigskip
\bigskip

\Large\textbf{Fig.~3}

\clearpage

\normalsize
\begin{figure}[h]
\caption{Details of skyrmions solutions for $\eta=0.4$ 
and $b=0.05$ at various temperatures. 
(See, Eqs.~(\ref{equM}),(\ref{equTheta}) in Methods section.)
\textbf{a},
Variation of the modulus of the magnetisation
as function of radius at different temperatures.
\textbf{b},
Variation of the angle of rotation, $\theta$, as function
of radius at different temperatures.
At low temperatures the magnetization rotates from angle zero 
in the center to angles $ \theta(R) < \pi$ at the cell boundaries.
The angles $\theta(R)$ depend on temperature. 
The radius is given in units of the twisting lengths $1/q_0$.
\textbf{c}, 
Evolution of the radius of the skyrmion cell
up to the transition into the paramagnetic state.
In the transition region both the radius of the skyrmion cell $R$ 
and the angles $ \theta(R)$ grow unlimitedly
by approaching the critical temperature $T_S$.
\textbf{d}, 
Corresponding evolution for the amplitude 
magnetization in the center $m(0)$.
The modulus gradually decreases to zero as the temperature 
approaches the critical value $T_S$ 
while the radius of the skyrmions diverges.
By this process the single skyrmions forming a dense stable 
condensate below $T_S$ can transform into isolated skyrmions.
Note that the characteristic twisting length in radial direction,
which is given by the slope $\theta(\rho)$ in panel \textbf{b},
does not change with changing temperature.
The core region  of the skyrmions with sizeable magnetization has
essentially a fixed diameter, while only a tail with diverging
radius and rapidly vanishing local magnetization evolves 
near $T_S$.
This diameter is given by the ratio $D/A$, which 
also fixes the characteristic size of chiral 
fluctuations in the paramagnetic phase above $T_S$.
}
\label{Fig3}
\end{figure}

\clearpage

\centerline{\includegraphics[width=16cm]{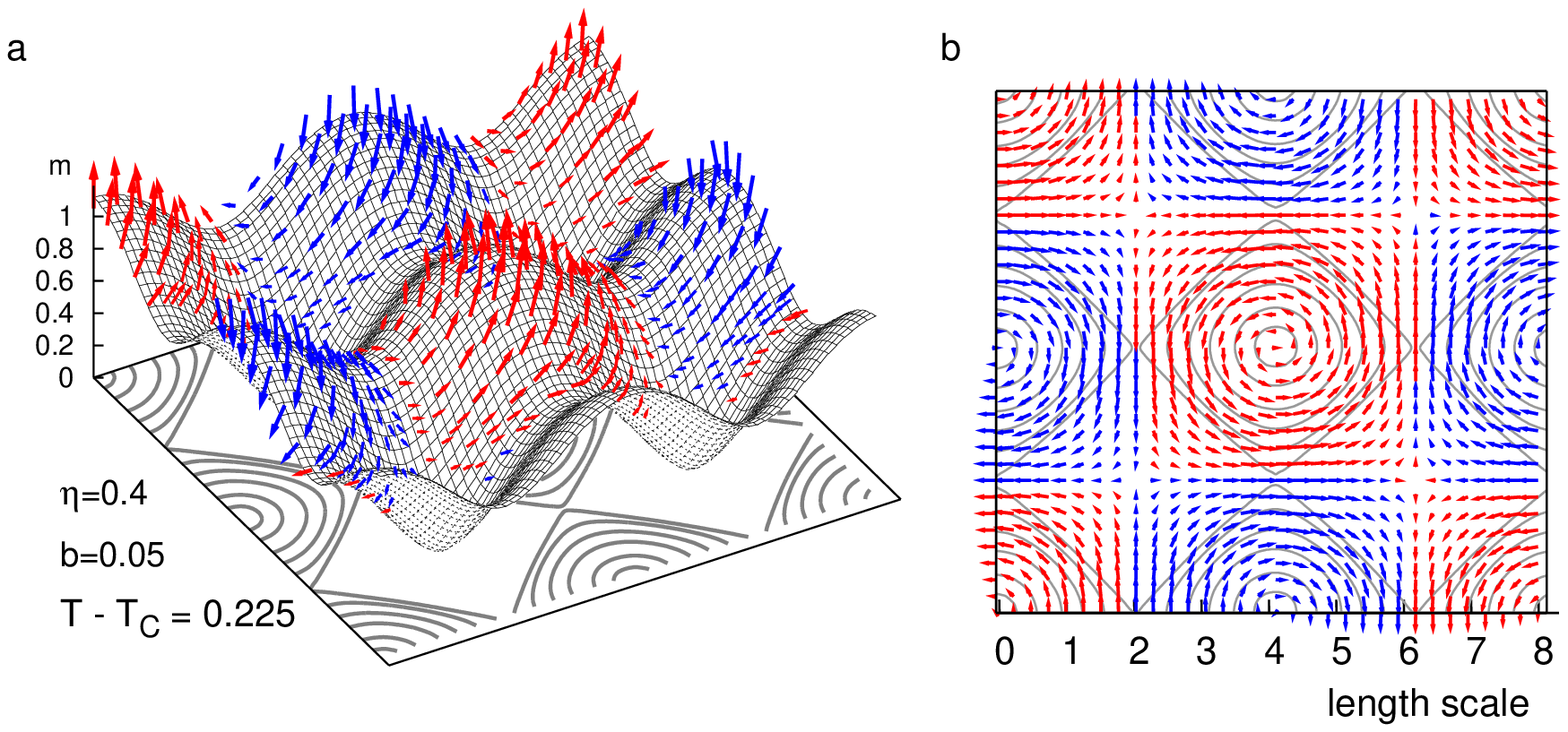}}

\bigskip
\bigskip

\Large\textbf{Fig.~4}

\clearpage

\normalsize
\begin{figure}[h]
\caption{
Structure of a two-dimensional skyrmion lattice,
derived as minimum energy solution 
for the model Eqn.\,(\ref{energy})
with Dzyaloshinskii-Moriya interactions, Eqn.\,(\ref{DzMo}).
The ground-state has four-fold symmetric lattice structure,
and net magnetization direction of the skyrmion 
cores is staggered. 
Thus, the periodicity of this square lattice 
is doubled as a 2$\times$2 antiferromagnetic structure. 
\textbf{a}, overview showing the modulus $m$ and the corresponding
magnetization vectors. 
\textbf{b}, Projection of the magnetization 
vectors into the base-plane.
Units shown in \textbf{b} are the twisting length $1/q_0$.
Red/blue signals local magnetization direction 
with positive/negative components 
arbitrarily taken perpendicular to the two-dimensional plane.
For the core-regions of the skyrmions,
lines of constant modulus, $m=$const,
are shown as contour lines in the base-plane.
This result applies to thin magnetic films with 
broken inversion symmetry
made from isotropic or cubic (crystallographic class T) 
metallic materials.
The magnetization direction rotates 
around the radial directions going outward from 
the skyrmion center.
A three-dimensional analogue of this dense texture
composed of an amorphous arrangement of cylindrical skyrmion 
strings is consistent with recent experiments in the cubic B20 metals 
FeGe and MnSi \cite{lebe89,lebe93,pfle04} as discussed in the text.
}
\label{Fig4}
\end{figure}


\begin{thebibliography}{99}
\bibitem{dipo}
Note that dipolar interactions  do not commute with ferromagnetic and antiferromagnetic order. 
Further, dipolar interactions are larger the larger the ordered magnetic moment. 
This suggests that $\eta<1$ even in the limit $q\to0$. While classical dipolar interactions 
may suppress the formation of certain skyrmions they favour the skyrmions relevant 
for isotropic and cubic ferromagnets, as shown in Figure~4, which have a structure that 
closes the magnetic flux (with $\phi=\pm \pi$, see Methods section).
\end{thebibliography}
\end{document}